\documentclass[12pt]{article}

\input{epsf.tex}

\newcommand{\E}{{\cal{E}}}

\renewcommand{\a}{\alpha}

\newcommand{\be}{\begin{equation}}
\newcommand{\ee}{\end{equation}}
\newcommand{\bea}{\begin{eqnarray}}
\newcommand{\eea}{\end{eqnarray}}
\def\J#1#2#3#4{#1 {\it #2} {\bf #3} #4}
\def\PTP{\it Prog. Theor. Phys.}

\def\PRD{{\it Phys. Rev.} D}
\def\PR{\it Phys. Rev.}

\def\JMP{\it J. Math. Phys.}
\def\CQG{\it Class. Quantum Grav.}

\def\PLA{\it Phys. Lett. A}

\begin{document}
\title{\bf\Large Physical interpretation of NUT solution}
\author{V~S~Manko\dag{} and E~Ruiz\ddag}
\date{}
\maketitle

\vspace{-1cm}

\begin{center}
\dag Departamento de F\'\i sica,\\ Centro de Investigaci\'on y de
Estudios Avanzados del IPN,\\ A.P. 14-740, 07000 M\'exico D.F.,
Mexico\\ \ddag Area de F\'\i sica Te\'orica, Universidad de
Salamanca,\\ 37008 Salamanca, Spain
\medskip
\end{center}

\vspace{.1cm}

\begin{abstract}
\noindent We show that the well--known NUT solution can be
correctly interpreted as describing the exterior field of two
counter--rotating semi--infinite sources possessing negative
masses and infinite angular momenta which are attached to the
poles of a static finite rod of positive mass.
\end{abstract}

\noindent \hspace{1cm} PACS numbers: 04.20.Jb

\vspace{6cm}


\newpage

\section{Introduction}

\noindent The Newman--Tamburino--Unti (NUT) spacetime is defined
by the metric \cite{NTU,KSt} \bea
ds^2&=&f^{-1}dr^2+(r^2+\nu^2)(d\vartheta^2+\sin^2\vartheta
d\varphi^2)-f(dt-2\nu\cos\theta d\varphi)^2, \nonumber\\
f&=&(r^2-2mr-\nu^2)/(r^2+\nu^2), \label{m_nut} \eea where the
parameter $m$ represents the mass of the source, and $\nu$ is the
so--called NUT parameter which was given in the literature several
other names, the most commonly known being probably
``gravomagnetic monopole''.

If $\nu\neq 0$, the metric (\ref{m_nut}) is stationary,
axisymmetric, but not globally asymptotically flat because it has
two singularities on the symmetry axis defined by $\vartheta=0$
and $\vartheta=\pi$ (in the original paper \cite{NTU} the
singularity $\vartheta=0$ is removed by performing a trivial
redefinition of the $dtd\varphi$ term in (\ref{m_nut})).

There are two basic approaches to the interpretation of the NUT
solution. The first one employed by Misner \cite{Mis} aims at
eliminating singular regions at the expense of introducing the
``periodic'' time and, consequently, the  $S^3$ topology for the
hypersurfaces $r={\rm const}$; it will not be discussed in our
paper. The second approach was used by Bonnor \cite{Bon} who
assumed the genuine character of the singularity $\vartheta=\pi$
of \cite{NTU} and interpreted it as a ``massless source of angular
momentum''. It was pointed out by Israel \cite{Isr} that the
Bonnor singularity is not a simple line source such as, e.g., a
massless strut appearing in the systems of two static particles.

Guided by the above Israel's remark, we have reexamined the work
\cite{Bon} and found out that Bonnor's interpretation suffers the
following two fundamental defects: (i) the semi--infinite
singularity of paper \cite{Bon} is not really massless, and (ii)
its angular momentum calculated, like its mass, via Komar
integrals \cite{Kom} has an infinitely large magnitude, thus
contradicting the well--known fact that the total angular momentum
of the NUT solution is equal to zero.

In the present letter we will show that the presence of two
semi--infinite counter--rotating singular regions endowed with
negative masses and infinitely large angular momenta is imperative
for providing a correct physical interpretation of the NUT metric.

\section{Behavior of NUT solution on the symmetry axis and
Komar quantities}

\noindent It is advantageous to analyze the properties of NUT
solution in the standard Weyl--Papapetrou cylindrical coordinates,
in terms of which the general stationary axisymmetric line element
has the form \be d s^2=f^{-1}[e^{2\gamma}(d\rho^2+d z^2)+\rho^2
d\varphi^2]-f(d t-\omega d\varphi)^2, \label{Papa} \ee the metric
coefficients $f$, $\gamma$, $\omega$ being functions of $\rho$ and
$z$ only.

In the NUT case the expressions for $f$, $\gamma$ and $\omega$
were given by Gautreau and Hoffman \cite{GHo}, but below these are
presented in a slightly different form, accompanied by the complex
Ernst potential $\E$ \cite{Ern} defining the NUT solution: \bea
&&\E=f+i\Omega=\frac{\a x-m-i\nu}{\a x+m+i\nu}, \quad
\Omega=-\frac{2\a\nu x}{(\a x+m)^2+\nu^2}, \nonumber\\
&&f=\frac{\a^2(x^2-1)}{(\a x+m)^2+\nu^2}, \quad
e^{2\gamma}=\frac{x^2-1}{x^2-y^2}, \quad \omega=2\nu y+C,\nonumber\\
&&x=\frac{1}{2\a}(r_++r_-), \quad y=\frac{1}{2\a}(r_+-r_-),
\nonumber\\ &&r_\pm=\sqrt{\rho^2+(z\pm\a)^2}, \quad
\a=\sqrt{m^2+\nu^2}, \label{mfun} \eea where $C$ is an arbitrary
real constant. In the paper \cite{NTU} $C$ was given the value
$-2\nu$; our choice of $C$ assumed throughout this letter and
leading to (\ref{m_nut}) is $C=0$. Mention that formulae
(\ref{mfun}) are a particular vacuum specialization of the more
general relations obtained in \cite{AGM} for the
Demia\'nski--Newman electrovac metric \cite{DNe}.

\begin{figure}[htb]
\centerline{\epsfysize=80mm\epsffile{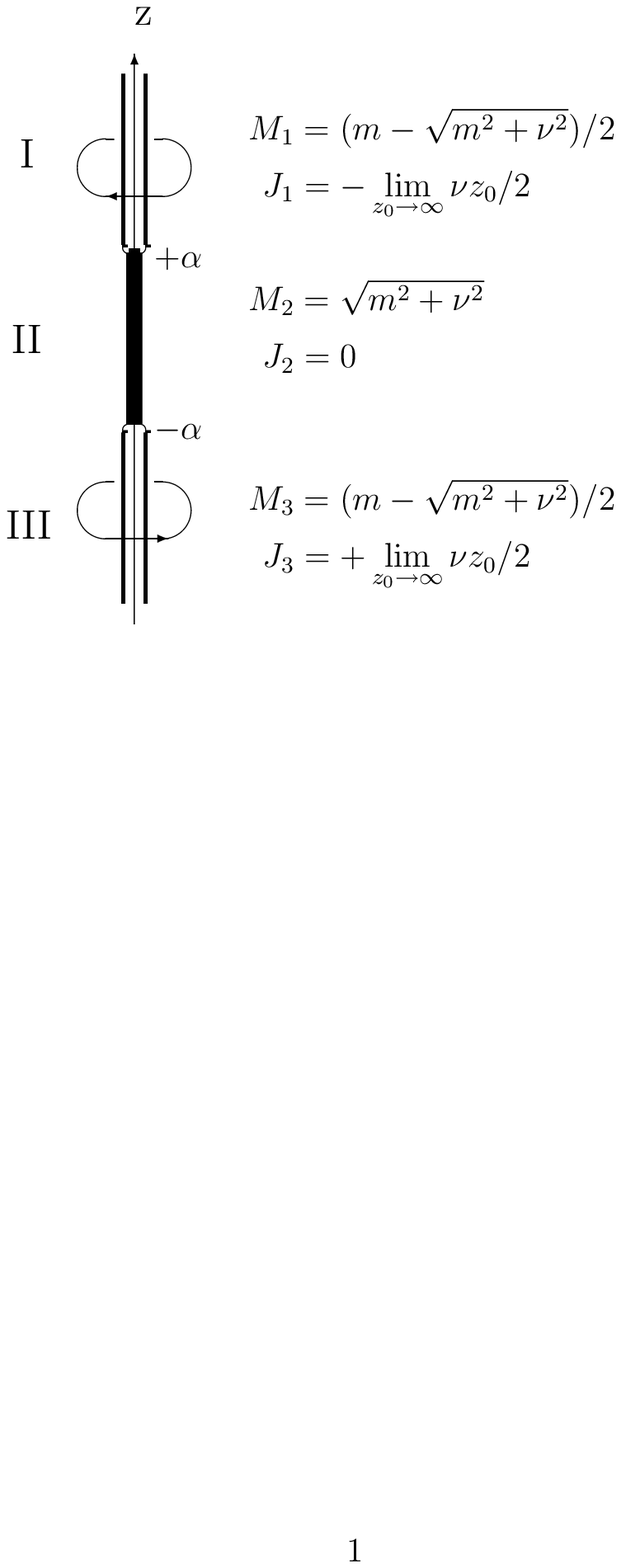}} \caption{Sources of
the NUT spacetime: two semi--infinite counter--rotating rods of
negative masses (regions~I and III) and a finite static rod of
positive mass (region~II).}
\end{figure}

The singularities of NUT solution lie on the symmetry $z$--axis
where one should distinguish the following three regions (see
figure~1): \be z>\a \hspace{0.2cm}({\rm region\,\,I}), \quad
|z|<\a \hspace{0.2cm}({\rm region\,\,II}), \quad z<-\a
\hspace{0.2cm}({\rm region\,\,III}). \label{regions} \ee In what
follows we shall consider the axis values of $f$, $\Omega$ and
$\omega$ under the above choice $C=0$, which are needed for the
calculation of the Komar masses and angular momenta.

{\it Region I}. The upper part of the symmetry axis is defined by
$\rho=0$, $z>\a$, or $x=z/\a$, $y=1$. Substituting the latter
values of $x$ and $y$ into (\ref{mfun}) we get \be
f=\frac{z^2-m^2-\nu^2}{(z+m)^2+\nu^2}, \quad \Omega=-\frac{2\nu
z}{(z+m)^2+\nu^2}, \quad \omega=2\nu. \label{mfun1} \ee

{\it Region II}. The second (intermediate) region is defined by
$\rho=0$, $|z|<\a$, or $x=1$, $y=z/\a$, and the corresponding axis
values of $f$, $\Omega$ and $\omega$ are \be f=0, \quad
\Omega=-\frac{\nu}{\a+m}, \quad \omega=2\nu z/\a. \label{mfun2}
\ee (Note that $\Omega$ assumes a constant value here.)

{\it Region III}. On the lower part of the symmetry axis
($\rho=0$, $z<-\a$, or $x=-z/\a$, $y=-1$) we have \be
f=\frac{z^2-\a^2}{(z-m)^2+\nu^2}, \quad \Omega=\frac{2\nu
z}{(z-m)^2+\nu^2}, \quad \omega=-2\nu. \label{mfun3} \ee

The non--vanishing constant values of $\omega$ on the first and
third segments of the axis signify that the NUT solution has two
semi--infinite singularities there. Although the upper singularity
can be easily removed by adding to $\omega$ in (\ref{mfun}) the
constant value $-2\nu$ (this was done by Bonnor in \cite{Bon}; of
course, instead of the upper singularity one may always eliminate
the lower one by adding to $\omega$ the constant value $2\nu$), we
shall retain both singularities since only in that way it is
possible to achieve a physically (and mathematically)
non--contradictory interpretation of the NUT spacetime.

We shall now calculate the masses $M_i$ and angular momenta $J_i$
($i=1,2,3$) of the regions I--III using the Komar integrals
\cite{Kom} in the form \cite{Tom,MRS} \bea
&&M=\lim\limits_{\rho\to 0} \frac{1}{4}\int_{z_1}^{z_2}[\rho({\rm
ln}f)_{,\rho} -\omega\Omega_{,z}]dz, \nonumber\\
&&J=-\lim\limits_{\rho\to 0}
\frac{1}{8}\int_{z_1}^{z_2}[2\omega-2\omega\rho({\rm ln}f)_{,\rho}
+(\rho^2f^{-2}+\omega^2)\Omega_{,z}]dz, \label{Komar} \eea where
$z_1$ and $z_2$ are two arbitrary points on the $z$--axis. In the
case of the region~I we have to assign to $z_1$ and $z_2$ the
values $\a$ and $+\infty$, respectively; moreover, $z_1=-\a$,
$z_2=\a$ (region~II) and $z_1=-\infty$, $z_2=-\a$ (region~III).

During the evaluation of the integrals (\ref{Komar}) one has to
take into account that \be \lim\limits_{\rho\to 0}\rho({\rm
ln}f)_{,\rho}=\left\{
\begin{array}{ll} 0, \quad {\rm regions\,\, I\,\, and\,\, III}\\
2, \quad {\rm region\,\, II}\\
\end{array}\right. \label{limro} \ee and also that $\Omega_{,z}=0$ in the
region~II.

A trivial integration then yields the following result for $M_i$:
\be M_1=M_3=\frac{1}{2}(m-\sqrt{m^2+\nu^2}), \quad
M_2=\sqrt{m^2+\nu^2}, \label{mass_Kom} \ee whence it follows that
for all $\nu\neq 0$ the masses of the singular sources are equal
and negative.

Turning now to the angular momenta, it is easy to see that
$J_2=0$, i.e., the intermediate rod is static. On the other hand,
the integrals $J_1$ and $J_3$ are divergent due to the presence of
the constant term $2\omega$ in the integrands. Therefore, in order
to have a better idea about the angular momenta of the
singularities, we must calculate $J_1$ (and $J_3$) for some point
$z=z_0$ (and $z=-z_0$), $z_0>\a$, of the symmetry axis, tending
then $z_0$ to infinity. We obtain \bea J_1(z_0)&=&-\frac{1}{8}
\int_{\a}^{z_0}(2\omega+\omega^2\Omega_{,z})\Big|_{\rho=0}dz
\nonumber\\ &=&-\frac{1}{2}\nu(z_0-m)
+\frac{\nu^3z_0}{(z_0+m)^2+\nu^2}=-J_3(z_0). \label{mom_Kom} \eea

We thus infer that the angular momenta of the two singular
semi--infinite sources are antiparallel and equal in absolute
values. In the limit $z_0\to\infty$, $J_1$ and $J_3$ tend to
infinity as $\mp z_0\nu/2$; however, the total angular momentum of
NUT solution is equal to zero identically.

\section{Discussion}

The model of two counter--rotating sources is the only possibility
for the NUT solution to be stationary and possess zero total
angular momentum; it fixes the choice of the additive constant in
the expression for the metric function $\omega$. In the original
NUT metric considered by Bonnor, for instance, the latter constant
has the value $-2\nu$, leading to $\omega=2\nu(y-1)$ and to the
regular upper part of the symmetry axis with $J_1=0$; besides, the
rod $\rho=0$, $|z|<\a$ acquires the angular momentum $J_2=-\nu\a$.
However, the corresponding angular momentum of the semi--infinite
singularity becomes an infinitely large quantity
$J_3=\lim_{z_0\to\infty}\nu z_0$, and the mass $M_3$ takes the
value $m-\sqrt{m^2+\nu^2}<0$, thus invalidating Bonnor's
interpretation.

The parameter $m$ does not affect qualitatively our interpretation
of the NUT solution because for any value of $m$ and a
non--vanishing $\nu$, two rotating regions of negative mass and
one static rod of positive mass will be always present. We
anticipate that negative masses of the singular sources could have
direct relation with the formation of a specific causal structure
of NUT metric involving closed timelike curves, but further
investigation is needed to lend support to this hypothesis.

\vspace{1cm}

\noindent{\bf Acknowledgements}

\vspace{.5cm}

We are grateful to Jes\'us Mart\'\i n for many interesting
discussions. One of the authors (VSM) would like to thank the
Department of Fundamental Physics of the Salamanca University for
its kind hospitality and financial support of his visit. This work
was also supported by Research Project BFM2003--02121 from
Ministerio de Ciencia y Tecnolog\'\i a of Spain and by CONACyT of
Mexico.


\newpage

\begin{thebibliography}{99}

\bibitem{NTU} Newman~E, Tamburino~L and Unti~T \J{1963}{\JMP}{4}{915}

\bibitem{KSt} Kramer~D, Stephani~H, MacCallum~M~A~N and Herlt~E
1980 {\it Exact Solutions of Einstein's Field Equations}
(Cambridge: Cambridge University Press)

\bibitem{Mis} Misner~C~W 1967 {\it Relativity Theory and
Astrophysics} ed J Ehlers (Amer. Math. Soc., Providence, Rhode
Island) Vol.~1, p.~160

\bibitem{Bon} Bonnor~W~B 1969 {\it Proc. Camb. Phil. Soc.}
{\bf 66} 145

\bibitem{Isr} Israel~W \J{1977}{\PRD}{15}{935}

\bibitem{Kom} Komar~A \J{1959}{\PR}{113}{934}

\bibitem{GHo} Gautreau~R and Hoffman~R~B \J{1972}{\PLA}{39}{75}

\bibitem{Ern} Ernst~F~J \J{1968}{\PR}{167}{1175}

\bibitem{AGM} Aguilar--S\'anchez~J~A, Garc\'\i a~A~A and Manko~V~S
2001 {\it Gravit. Cosmology} {\bf 7} 149

\bibitem{DNe} Demia\'nski~M and Newman~E~T 1966 {\it
Bull. Acad. Polon. Sci. Ser. Math. Astron. Phys.} {\bf 14} 653

\bibitem{Tom} Tomimatsu~A \J{1983}{\PTP}{70}{385}

\bibitem{MRS} Manko~V~S, Ruiz~E and Sanabria--G\'omez~J~D
\J{2000}{\CQG}{17}{3881}

\end{thebibliography}
\end{document}